\documentclass[10pt]{article}

\usepackage[
  a4paper,
  top=0.65in, bottom=1.0in, left=0.5in, right=0.5in,
  columnsep=0.41in
]{geometry}

\usepackage{fontspec}
\defaultfontfeatures{Ligatures=TeX}
\newfontfamily\headingfont{Satoshi-Medium}[
  Path = fonts/, Extension = .otf,
  UprightFont = Satoshi-Medium, ItalicFont = Satoshi-MediumItalic
]
\newfontfamily\accentfont{MartianMono-Light75}[
  Path = fonts/, Extension = .ttf, UprightFont = MartianMono-Light75,
  LetterSpace = 4.0
]

\usepackage{xcolor}
\definecolor{alabaster}{HTML}{F7F3ED}
\definecolor{midnightviolet}{HTML}{423759}
\definecolor{darkmoss}{HTML}{314942}
\definecolor{sunburst}{HTML}{FFA632}
\definecolor{flame}{HTML}{FF8633}
\definecolor{tablerule}{gray}{0.82}

\usepackage{amsmath}
\usepackage{mathastext}
\AtBeginDocument{%
  \DeclareMathSymbol{0}{\mathalpha}{mtletterfont}{`0}%
  \DeclareMathSymbol{1}{\mathalpha}{mtletterfont}{`1}%
  \DeclareMathSymbol{2}{\mathalpha}{mtletterfont}{`2}%
  \DeclareMathSymbol{3}{\mathalpha}{mtletterfont}{`3}%
  \DeclareMathSymbol{4}{\mathalpha}{mtletterfont}{`4}%
  \DeclareMathSymbol{5}{\mathalpha}{mtletterfont}{`5}%
  \DeclareMathSymbol{6}{\mathalpha}{mtletterfont}{`6}%
  \DeclareMathSymbol{7}{\mathalpha}{mtletterfont}{`7}%
  \DeclareMathSymbol{8}{\mathalpha}{mtletterfont}{`8}%
  \DeclareMathSymbol{9}{\mathalpha}{mtletterfont}{`9}%
  \DeclareMathSymbol{\times}{\mathbin}{mtletterfont}{`×}%
}
\newcommand{\mhyphen}{\mbox{-}\kern-0.05em}
\usepackage{stfloats}
\usepackage{graphicx}
\usepackage{tikz}
\usepackage{eso-pic}
\usepackage{tcolorbox}
\tcbuselibrary{skins}
\usepackage{titlesec}
\usepackage{ragged2e}
\usepackage{enumitem}
\usepackage{microtype}
\usepackage{tabularx}
\newcolumntype{L}{>{\raggedright\arraybackslash}X}
\newcolumntype{P}[1]{>{\raggedright\arraybackslash}p{#1}}
\usepackage{colortbl}
\usepackage[colorlinks=true, linkcolor=midnightviolet, urlcolor=midnightviolet, citecolor=midnightviolet]{hyperref}

\renewcommand{\normalsize}{\fontsize{8}{12}\selectfont}
\AtBeginDocument{\normalsize}
\titleformat{\section}
  {\color{black}\accentfont\fontsize{10}{14}\selectfont}
  {}{0pt}{}
\titlespacing*{\section}{0pt}{1.2em}{0.5em}
\let\piosectionorig\section
\renewcommand{\section}[1]{\piosectionorig{\MakeUppercase{#1}}}

\titleformat{\subsection}
  {\color{black}\headingfont\fontsize{9}{12}\selectfont}
  {}{0pt}{}
\titlespacing*{\subsection}{0pt}{0.9em}{0.3em}

\newlength{\piofooterheight}
\newlength{\piomastheadpad}

\AddToShipoutPictureBG{%
  \begin{tikzpicture}[remember picture, overlay, x=1pt, y=1pt]
    \fill[alabaster] (current page.south west) rectangle
      ([yshift=\piofooterheight]current page.south east);
    \node[anchor=south west, inner sep=0pt]
      at ([xshift=0.5in, yshift=0.5\piofooterheight-3pt]current page.south west)
      {\includegraphics[height=6pt]{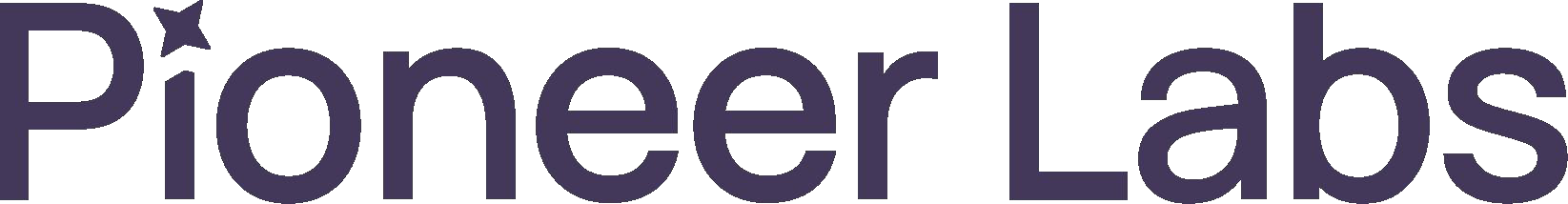}};
    \node[anchor=south east, inner sep=0pt, font=\accentfont\fontsize{5}{6}\selectfont, text=black]
      at ([xshift=-0.5in, yshift=0.5\piofooterheight-2.2pt]current page.south east)
      {\ifnum\value{page}<10 0\fi\thepage};
    \ifnum\value{page}=1
      \coordinate (piombY) at ([yshift=-\piomastheadpad]mastheadbottom);
      \fill[alabaster] (current page.north west) rectangle (piombY -| current page.east);
    \fi
  \end{tikzpicture}%
}

\newtcolorbox{piofigurebox}{
  enhanced,
  colback = white, colframe = midnightviolet,
  boxrule = 0.5pt, arc = 0pt, outer arc = 0pt,
  left = 0pt, right = 0pt, top = 0pt, bottom = 0pt, boxsep = 0pt,
  overlay = {
    \draw[midnightviolet, line width=4pt]
      ([xshift=2pt, yshift=-0.5pt]frame.north west) --
      ([xshift=2pt, yshift=-85pt]frame.north west);
  }
}
\newtcolorbox{piocaptionbox}{
  colback = alabaster, colframe = alabaster, boxrule = 0pt, arc = 0pt,
  left = 9.2pt, right = 10pt, top = 17.3pt, bottom = 5.6pt,
  width = \linewidth,
}
\newcounter{piofig}
\newcommand{\piofigure}[3]{%
  \begin{figure*}[t]
    \begin{piofigurebox}
      \noindent{\color{midnightviolet}\vrule width 1pt}%
      \begin{minipage}[t]{\dimexpr\linewidth-1pt\relax}
      \par\vspace{12pt}
      \hspace{16pt}%
      \begin{minipage}[t]{\dimexpr\linewidth-28pt\relax}
        \centering
        \includegraphics[width=\linewidth, height=0.4\textheight, keepaspectratio]{#1}
      \end{minipage}\par
      \vspace{9pt}
      \begin{piocaptionbox}
        \begingroup\justifying\setlength{\parindent}{0pt}\noindent
        {\accentfont\fontsize{7}{9}\selectfont
          \refstepcounter{piofig}Figure~\thepiofig: #2}%
        \ifx&#3&\else\par\vspace{3pt}{\fontsize{7}{9}\selectfont #3}\fi
        \endgroup
      \end{piocaptionbox}
      \end{minipage}%
    \end{piofigurebox}
  \end{figure*}%
}
\newcommand{\piofigureh}[4]{%
  \begin{figure*}[t]
    \begin{piofigurebox}
      \noindent{\color{midnightviolet}\vrule width 1pt}%
      \begin{minipage}[t]{\dimexpr\linewidth-1pt\relax}
      \par\vspace{12pt}
      \hspace{16pt}%
      \begin{minipage}[t]{\dimexpr\linewidth-28pt\relax}
        \centering
        \includegraphics[width=\linewidth, height=#2\textheight, keepaspectratio]{#1}
      \end{minipage}\par
      \vspace{9pt}
      \begin{piocaptionbox}
        \begingroup\justifying\setlength{\parindent}{0pt}\noindent
        {\accentfont\fontsize{7}{9}\selectfont
          \refstepcounter{piofig}Figure~\thepiofig: #3}%
        \ifx&#4&\else\par\vspace{3pt}{\fontsize{7}{9}\selectfont #4}\fi
        \endgroup
      \end{piocaptionbox}
      \end{minipage}%
    \end{piofigurebox}
  \end{figure*}%
}
\newcommand{\piofigurehcol}[4]{%
  \begin{figure}[t]
    \begin{piofigurebox}
      \noindent{\color{midnightviolet}\vrule width 1pt}%
      \begin{minipage}[t]{\dimexpr\linewidth-1pt\relax}
      \par\vspace{12pt}
      \hspace{16pt}%
      \begin{minipage}[t]{\dimexpr\linewidth-28pt\relax}
        \centering
        \includegraphics[width=\linewidth, height=#2\textheight, keepaspectratio]{#1}
      \end{minipage}\par
      \vspace{9pt}
      \begin{piocaptionbox}
        \begingroup\justifying\setlength{\parindent}{0pt}\noindent
        {\accentfont\fontsize{7}{9}\selectfont
          \refstepcounter{piofig}Figure~\thepiofig: #3}%
        \ifx&#4&\else\par\vspace{3pt}{\fontsize{7}{9}\selectfont #4}\fi
        \endgroup
      \end{piocaptionbox}
      \end{minipage}%
    \end{piofigurebox}
  \end{figure}%
}
\newcommand{\piofigurelabeled}[4]{%
  \begin{figure*}[t]
    \begin{piofigurebox}
      \noindent{\color{midnightviolet}\vrule width 1pt}%
      \begin{minipage}[t]{\dimexpr\linewidth-1pt\relax}
      \par\vspace{12pt}
      \hspace{16pt}%
      \begin{minipage}[t]{\dimexpr\linewidth-28pt\relax}
        \centering
        \includegraphics[width=\linewidth, height=0.4\textheight, keepaspectratio]{#1}
      \end{minipage}\par
      \vspace{9pt}
      \begin{piocaptionbox}
        \begingroup\justifying\setlength{\parindent}{0pt}\noindent
        {\accentfont\fontsize{7}{9}\selectfont Figure~#2: #3}%
        \ifx&#4&\else\par\vspace{3pt}{\fontsize{7}{9}\selectfont #4}\fi
        \endgroup
      \end{piocaptionbox}
      \end{minipage}%
    \end{piofigurebox}
  \end{figure*}%
}
\newcommand{\piofigurelabeledinline}[4]{%
  \par\vspace{6pt}\noindent
  \begin{piofigurebox}
    \noindent{\color{midnightviolet}\vrule width 1pt}%
    \begin{minipage}[t]{\dimexpr\linewidth-1pt\relax}
    \par\vspace{8pt}
    \hspace{16pt}%
    \begin{minipage}[t]{\dimexpr\linewidth-28pt\relax}
      \centering
      \includegraphics[width=0.74\linewidth, height=0.20\textheight, keepaspectratio]{#1}
    \end{minipage}\par
    \vspace{7pt}
    \begin{piocaptionbox}
      \begingroup\justifying\setlength{\parindent}{0pt}\noindent
      {\accentfont\fontsize{7}{9}\selectfont Figure~#2: #3}%
      \ifx&#4&\else\par\vspace{3pt}{\fontsize{7}{9}\selectfont #4}\fi
      \endgroup
    \end{piocaptionbox}
    \end{minipage}%
  \end{piofigurebox}\par\vspace{6pt}
}

\newcounter{piotable}
\newcommand{\piotablecaption}[3][\textwidth]{%
  \refstepcounter{piotable}%
  \begin{tcolorbox}[
      colback=alabaster, colframe=alabaster, boxrule=0pt, arc=0pt,
      left=5.4pt, right=8pt, top=8.4pt, bottom=5.8pt,
      width=#1
    ]
    \justifying\setlength{\parindent}{0pt}\noindent\fontsize{8}{11}\selectfont
    {\accentfont\fontsize{8}{11}\selectfont Table~\thepiotable. #2}~#3
  \end{tcolorbox}\vspace{-0.55em}\vspace{-2.6pt}%
}
\newcommand{\piotablecaptionnonum}[3][\textwidth]{%
  \begin{tcolorbox}[
      colback=alabaster, colframe=alabaster, boxrule=0pt, arc=0pt,
      left=5.4pt, right=8pt, top=8.4pt, bottom=5.8pt,
      width=#1
    ]
    \justifying\setlength{\parindent}{0pt}\noindent\fontsize{8}{11}\selectfont
    {\accentfont\fontsize{8}{11}\selectfont #2}~#3
  \end{tcolorbox}\vspace{-0.55em}\vspace{-2.6pt}%
}
\newcommand{\piohead}[1]{\cellcolor{midnightviolet}\textcolor{white}{#1}}
\arrayrulecolor{tablerule}

\newcommand{\piodivider}{\noindent{\color{midnightviolet}\rule{0.83\linewidth}{0.25pt}}\par}

\newenvironment{pioneerrefsnum}{%
  \begingroup\fontsize{7.5}{10.5}\selectfont
  \setlength{\parindent}{0pt}
  \begin{enumerate}[leftmargin=1.6em, itemindent=0pt, itemsep=0.22em, label=\arabic*.]
}{%
  \end{enumerate}\endgroup
}

\begin{document}

\setlength{\hoffset}{0.1in}
\AddToHook{shipout/before}{\ifnum\value{page}>1\setlength{\hoffset}{0pt}\fi}

\twocolumn[
  {\raggedright
  \begin{minipage}[t]{0.72\linewidth}
    \raggedright
    {\headingfont\fontsize{22pt}{26pt}\selectfont\color{midnightviolet}
      A Framework for Evaluating Forward Contamination Risk for bio-ISRU Microorganisms\par}
  \end{minipage}\hfill
  \begin{minipage}[t]{0.24\linewidth}
    \raggedleft\vspace{2pt}
    \includegraphics[width=0.95\linewidth]{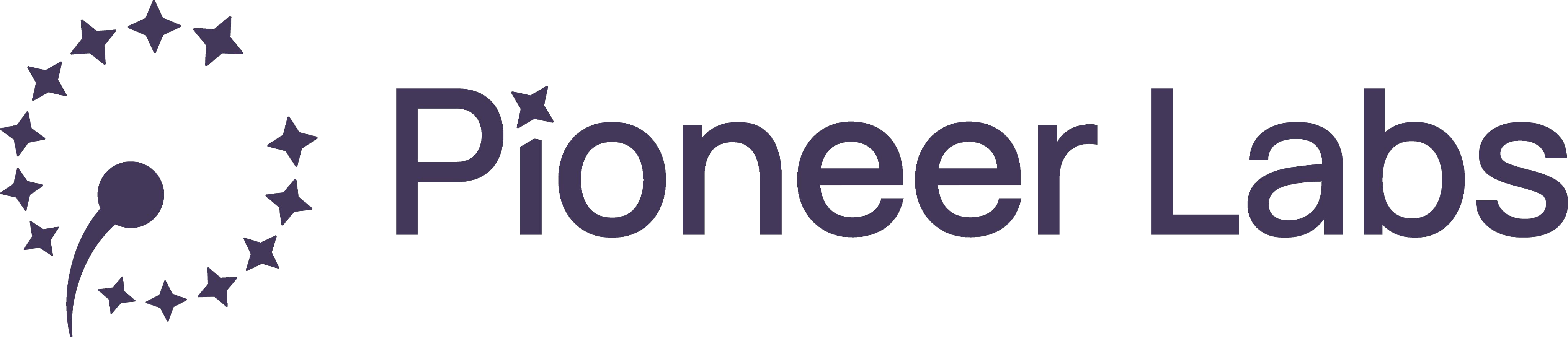}
  \end{minipage}

  \vspace{15pt}
  {\fontsize{10}{13}\selectfont September 9, 2026\par}
  \vspace{2.5pt}
  \piodivider
  \vspace{2.5pt}
  {\fontsize{10}{13}\selectfont\textbf{Lead Scientists}: Una Nattermann, Devon A. Stork\par}
  \vspace{2.5pt}
  \piodivider
  \vspace{2.5pt}
  {\fontsize{10}{13}\selectfont\textbf{Pioneer Labs}: Tom Pedersen, Nathan D. Hicks, Jordan E. Mancuso, Keren Isaev, Max G. Schubert,\\ Fatima R. Martin, Jonathan Liu, Harley Greene, Edward Sukarto, Una Nattermann, Devon A. Stork,\\ Erika A. DeBenedictis\par}
  \vspace{2.5pt}
  \piodivider
  \vspace{2.5pt}
  {\fontsize{10}{13}\selectfont\textbf{Affiliations}: All authors are affiliated with Pioneer Labs, Emeryville, CA 94608, USA.\par}
  \vspace{2.5pt}
  \piodivider
  \vspace{2.5pt}
  {\fontsize{10}{13}\selectfont\textbf{Contact}: \href{mailto:bioisru@pioneer-labs.org}{\textcolor{black}{bioisru@pioneer-labs.org}}\par}
  }%
  {\tikz[remember picture,overlay,x=1pt,y=1pt]\coordinate (mastheadbottom) at (0,0);}%

  \vspace{\piomastheadpad}\vspace{12.3pt}
  {\accentfont\fontsize{10}{14}\selectfont\color{black} ABSTRACT:}\par\vspace{4pt}
  \fontsize{8}{12}\selectfont
  NASA planetary protection policy seeks to avoid inadvertent forward
  contamination of celestial bodies and backward contamination of Earth
  during space exploration. For Mars missions, it has historically focused
  on sterilization of landers and rovers. Current policy provides no
  framework for assessing the forward contamination risk of missions that
  deliberately include living organisms and must not be sterilized, such as
  a human crew, agriculture, or biotechnology for manufacturing and life
  support. We propose the \textbf{PRIM} (\textbf{\underline{P}}ropagation
  \textbf{\underline{R}}estricted, \textbf{\underline{I}}nert on
  \textbf{\underline{M}}ars) framework to evaluate the forward-contamination
  risk posed by known organisms. Building on the biocontainment literature
  and probability-of-contamination model, PRIM bounds the ability of an
  organism released from a worst-case off-nominal event to sustain growth on
  the Martian surface at a probability under $P_c \leq 10^{\mhyphen 4}$ using
  independent single-stressor propagation assays. We apply the PRIM
  framework to two engineered Mars biological \textit{in situ} resource
  utilization chassis organisms, using low water activity and
  carbon starvation to qualify them for low forward contamination risk even
  at high bioburden, and outline how to extend the assays to phototrophs and
  anaerobes. PRIM offers a path to flight-qualify living organisms for Mars
  based on laboratory assays rather than bioburden alone. This enables
  biological life support infrastructure for a sustained presence on Mars
  while keeping forward contamination risk demonstrably low.
  \vspace{22pt}
]

\section{Introduction}

Human exploration of Mars will deliberately send terrestrial biology to the Red
Planet. Crewed missions will carry human-associated microbial communities that
cannot be sterilized, making the introduction of terrestrial organisms to Mars
inevitable\textsuperscript{1}.
Independent of the crew, a growing body of work proposes microbes as infrastructure
for microbial waste processing, life-support
loops\textsuperscript{2}, and
biological \textit{in situ} resource utilization (bio-ISRU) to convert local
regolith, water ice, and atmospheric CO\textsubscript{2} into building materials,
propellant, food, and
pharmaceuticals\textsuperscript{3-7}.
This includes worked mission concepts incorporating bio-ISRU as
an integral component that enables the human exploration of
Mars\textsuperscript{3,4,8}.
These applications can use well-characterized non-extremophile organisms
that are unable to tolerate the pressure, temperature, and energy scarcity of the
Mars environment.

Planetary protection exists to preserve the scientific integrity of Mars and is
grounded in Article IX of the 1967 Outer Space
Treaty\textsuperscript{9}, implemented
internationally through the COSPAR Policy on Planetary
Protection\textsuperscript{10} and, for
U.S. missions, through NASA's planetary-protection
requirements\textsuperscript{11}. The
underlying probability-of-contamination model, conventionally known as the
Coleman-Sagan
formula\textsuperscript{12-14},
expresses the probability of contamination as
\[
P_c = N_0 \cdot R \cdot P_S \cdot P_I \cdot P_R \cdot P_g
\]
where $N_0$ is the initial number of organisms on the spacecraft at launch, $R$ is
the reduction achieved by sterilization conditions, $P_S$ is the probability that an
organism reaches the surface, $P_I$ is the probability of impact, $P_R$ is the
probability of release into the environment, and $P_g$ is the probability that a
released organism grows. A mission complies with planetary protection if
$P_c \leq 10^{\mhyphen 4}$, the threshold in use since
Viking\textsuperscript{15,16}.

On Mars, acceptable bioburden thresholds are held to stricter limits across the
planet and especially for local environments on Mars thought capable of supporting
propagation, which are designated as ``Special
Regions''\textsuperscript{17,18}.
These areas are defined as those that could simultaneously reach a water activity
($a_w$) $\geq 0.5$ and a temperature $\geq \mhyphen 28^{\circ}\mathrm{C}$\textsuperscript{19}.
Current policy dictates that wherever and whenever those thresholds are
simultaneously met, it is theoretically possible for life to propagate (\textit{i.e.}\ $P_g$
is expected to be one) regardless of the organism carried. This policy is
implemented through reducing bioburden on outbound hardware to acceptably low
levels\textsuperscript{11}, especially
of extremotolerant spore formers that survive cleanroom environments. For example,
the NASA Standard Assay is the operational method for measuring spacecraft
bioburden\textsuperscript{11}, and it
measures $N_0$. This is a sensible approach to reduce the potential for forward
contamination when dealing with a heterogeneous, largely uncharacterized set of
environmental contaminants whose identity and physiology are unknown, and whose
$P_g$ cannot be known.

Moving forward, the stated goal of space agencies around the world is for a human
presence on Mars. A human gut and skin microbiome, or plant-based agriculture,
would be large, diverse, partly unculturable communities introduced at high titer.
Planetary protection policies will have to drastically adapt to accommodate crewed
surface operations. Therefore, there is considerable interest in updating planetary
protection policy to guide risk assessment for a new era of missions that include
humans, agriculture, and biotechnology that require deliberate transfer of living
organisms to other
planets\textsuperscript{20--23}.
For an organism sent deliberately, $P_g$ can be assessed
experimentally. Known, clonal, well-characterized chassis can be tested directly
under Mars Special Region water activities, temperatures, and energy sources, to a
defined limit of detection. We propose the \textbf{PRIM}
(\textbf{\underline{P}}ropagation \textbf{\underline{R}}estricted,
\textbf{\underline{I}}nert on \textbf{\underline{M}}ars) framework to conduct
measurements and establish $P_g$ for the most permissive areas of Mars,
flight-qualifying a specific organism or defined community up to a maximum $N_0$
population size where $P_c \leq 10^{\mhyphen 4}$. In what follows, we detail the PRIM
framework, state the framework's assumptions, apply it to two engineered bio-ISRU
chassis in a worked example, and outline how to adjust the framework for other
organisms, such as phototrophs and anaerobic perchlorate reducers.

\section{The PRIM Framework}

The PRIM framework is a set of laboratory experiments to measure an upper bound on
$P_g$ for a specific clonal organism or defined community by characterizing both
survival and propagation under Mars conditions. Survival assays
quantify the fraction of cells that remain viable after exposure to a Martian
stressor(s); propagation assays investigate whether any surviving cells can
reproduce under that stressor (\textbf{Materials and Methods}). The two assays
answer different planetary protection questions. A high kill rate is reassuring yet
insufficient; for example, releasing $10^{12}$ cells in which 99.9999\% of the
population dies still leaves $10^6$ survivors, which risks a forward contamination
event if any cells can propagate. Cell death is important, but the inability to
propagate is what truly prevents forward contamination. Evolution cannot proceed
without growth, so an organism that cannot propagate on the Martian surface cannot
evolve its way toward one that can.

\textbf{Propagation assays}. We seed a high-titer culture into a condition
permissive for growth in every respect except a single Martian stressor under test,
and we monitor for any sustained increase in population during an extended growth
experiment (\textbf{Materials and Methods}). If the organism of interest
propagates, cell density increases. If it does not propagate, we expect the cell
density to stay constant or decline over time. We terminate each propagation assay
with a survival assay to convert the endpoint into a viable-cell count. The
propagation result provides an upper bound for forward contamination risk: no
growth was observed down to the sensitivity of the terminal survival assay
measurement.

\textbf{Survival assays.} At the end of the propagation assay, we take the entire
culture that was inoculated to a high-titer and exposed to the Martian stressor and
quantify the number of remaining viable cells by colony counting on rich permissive
media (\textbf{Materials and Methods}). Plating on rich media after the stress
exposure gives every remaining viable cell its best possible chance to replicate.
The ratio of viable cells recovered to cells originally exposed is the survival
fraction for that stressor. When no viable cells are recovered, the survival
fraction is reported as the assay's 95\% confidence limit of detection. This style
of assay is also used to quantify survival fraction after one-off exposure to
Martian stressors, such as UVC or freeze-thaw.

\textbf{Assay limit of detection.} The limit of detection (LOD) of any single assay
is set by the number of cells that can be included in a single
experiment\textsuperscript{24}. However,
instead of reporting the LOD as $1/N$, where $N$ is the number of cells tested, we
report it as $3/N$ to reflect a 95\% confidence interval following the statistical
rule of three for zero observed events\textsuperscript{25}. For example, if
$10^{12}$ cells were seeded at the beginning of an experiment and, after the
stressor, no propagation events were observed, the assay's LOD is calculated as
$3 \times 10^{\mhyphen 12}$.

\textbf{Experimental tractability.} The $P_g$ values PRIM needs to reach are
extremely low. For example, in a 1~mL bioreactor with high cell density, $N_0$ is
$\sim 10^9$ cells. To achieve a $P_c \leq 10^{\mhyphen 4}$, you require a
$P_g \leq 10^{\mhyphen 13}$. As bioreactor size and number of cells within increase, it
becomes infeasible to use single benchtop experiments to assess $P_g$. Imagine a
1,000~L bioreactor, a common industrial scale volume for biomanufacturing on Earth,
growing cells at high density. This could have upwards of $10^{16}$ cells, requiring
a $P_g \leq 10^{\mhyphen 20}$ to achieve the minimum safe $P_c$ level. Instead of exposing
thousands of liters of high-density cells to a single Martian stressor, we suggest
decomposing the requirement across several independent assays and multiplying their
individually calculated $P_g$ limits of detection to obtain a combined $P_g$:
\[
P_g^{\,combined} = P_g^{\,stressor\ 1} \cdot P_g^{\,stressor\ 2}
\]

\textbf{Stressor independence.} A calculated $P_g^{\,combined}$ is only accurate if
the individual stressors act independently and block growth through non-interacting
physiological mechanisms. This is represented mathematically by an $\varepsilon$
term near one in
\[
P_g^{\,combined} = \varepsilon \cdot P_g^{\,stressor\ 1} \cdot P_g^{\,stressor\ 2}
\]
Any individual application of PRIM needs to choose individual stressors that
minimally interact, or else quantify the extent to which a smaller or larger $\varepsilon$
renders the independent $P_g^{\,combined}$ model inaccurate. This could be done
through literature analysis to estimate $\varepsilon$, or experiments that calculate
individual $P_g$ for each stressor, determine the combined $P_g$, and calculate
$\varepsilon$ directly. For feasibility reasons, such experiments would likely need
to be performed under less stressful versions of the same conditions, such that the
calculated $P_g^{\,combined}$ is above the assay LOD. This would unavoidably reduce
confidence in $\varepsilon$, but directly calculating it for the stressors in
question would be no different than reducing the two stressors to a single stressor.

Published tests of combined growth inhibitors are reassuring about the likely size
and direction of the $\varepsilon$ interaction term for properly chosen stressors.
Studies of salt, pH, and weak-acid combinations found no detectable interaction in
growth inhibition\textsuperscript{26,27},
and the interaction terms that growth-boundary models do require appear only close
to the growth/no-growth limit and act to make conditions more inhibitory than the
product predicts, not less\textsuperscript{28,29}.
In fact, combined extremes are likely more inhospitable than measurements of
individual stressors suggest\textsuperscript{30}.
For example, UV irradiation can intensify the bactericidal effects of
perchlorate\textsuperscript{29}.
Applications of PRIM where the combined $P_g$ clears the requirement by more than
the plausible magnitude of $\varepsilon$ is robust to merely approximate
independence.

\textbf{Relevant Martian stressors.} There are numerous biocidal stressors on Mars
that, each and in combination, cause cell death and reduced propagation. However,
for a stress condition to be appropriate for PRIM, it must both be a condition that
a released organism would necessarily be exposed to and be relatively independent of
the other stressor conditions chosen. In this way, while UVC is a strong biocidal
agent at the surface, released organisms may be sheltered from UVC beneath regolith
or other obstructions. Similarly, high salinity and low-temperature robustness often
co-occur on Earth and therefore cannot be considered independently within the PRIM
framework. Here, we specifically discuss two stressors that impact the propagation
and survival of heterotrophic non-extremophilic bacteria, and are expected to be
relatively non-interacting.

\begin{itemize}[leftmargin=1.2em, itemsep=0.3em]
  \item \textbf{Low water activity.} Liquid water on Mars, where it may exist
    transiently, is expected to be a low-water-activity brine (likely
    perchlorate- and chloride-rich) with a water activity near the
    $a_w \sim 0.5$ threshold that defines a Special Region. While some
    extremophiles might survive in these
    conditions\textsuperscript{31--33},
    low water activity is a generally effective biocidal agent and creates
    conditions near impossible for propagation. We currently do not consider
    the possibility of transient nanometer-thick liquid water films that may
    have water activity near 1\textsuperscript{34,35}.
  \item \textbf{Carbon/Energy starvation.} Our chassis are heterotrophs, and the
    Martian regolith and atmosphere do not provide ample sources of reduced
    carbon. This stressor is specific to the organism and concerns the
    chassis' metabolism.
\end{itemize}

\section{Worked Example of a Worst-Case Scenario with Two Mars Bio-ISRU Bacteria}

\textbf{Scenario assumptions}. Here, we consider two worst-case scenarios: 1) an
off-nominal crash landing of a bioreactor containing the seed culture onto the
Martian surface and 2) a nominal landing after which a high-density growth
bioreactor experiences an off-nominal catastrophic leak that bypasses hardware containment
systems and spills its contents onto the Martian surface (\textbf{Figure 1A}).
Referring back to the Coleman-Sagan formula, in the first scenario $N_0 \sim 10^9$
and $\sim 10^{12}$ in the second. In both scenarios, we consider $R=1$, $P_S=1$,
$P_I=1$, and $P_R=1$, meaning we can simplify the formula to:
\[
P_c = N_0 \cdot P_g
\]
The number of organisms ($N_0$) that can be sent is limited by the probability that
those released organisms can grow ($P_g$), with the goal of that probability of
contamination ($P_c$) staying $\leq 10^{\mhyphen 4}$ (\textbf{Figure 1B}). Then, for the
first scenario where $N_0 \sim 10^9$, $P_g$ must be $\leq 10^{\mhyphen 13}$; and for the
second scenario where $N_0 \sim 10^{12}$, $P_g$ must be $\leq 10^{\mhyphen 16}$. In the
following worked example, we expand on a nominal landing and operation followed by a
catastrophic leak as the more difficult of the two scenarios. All together, the PRIM
framework quantifies the probability of contamination for our bio-ISRU strains where
the volume of the proposed bioreactor provides orders of magnitude separation from
the $P_c = 10^{\mhyphen 4}$.

\piofigurehcol{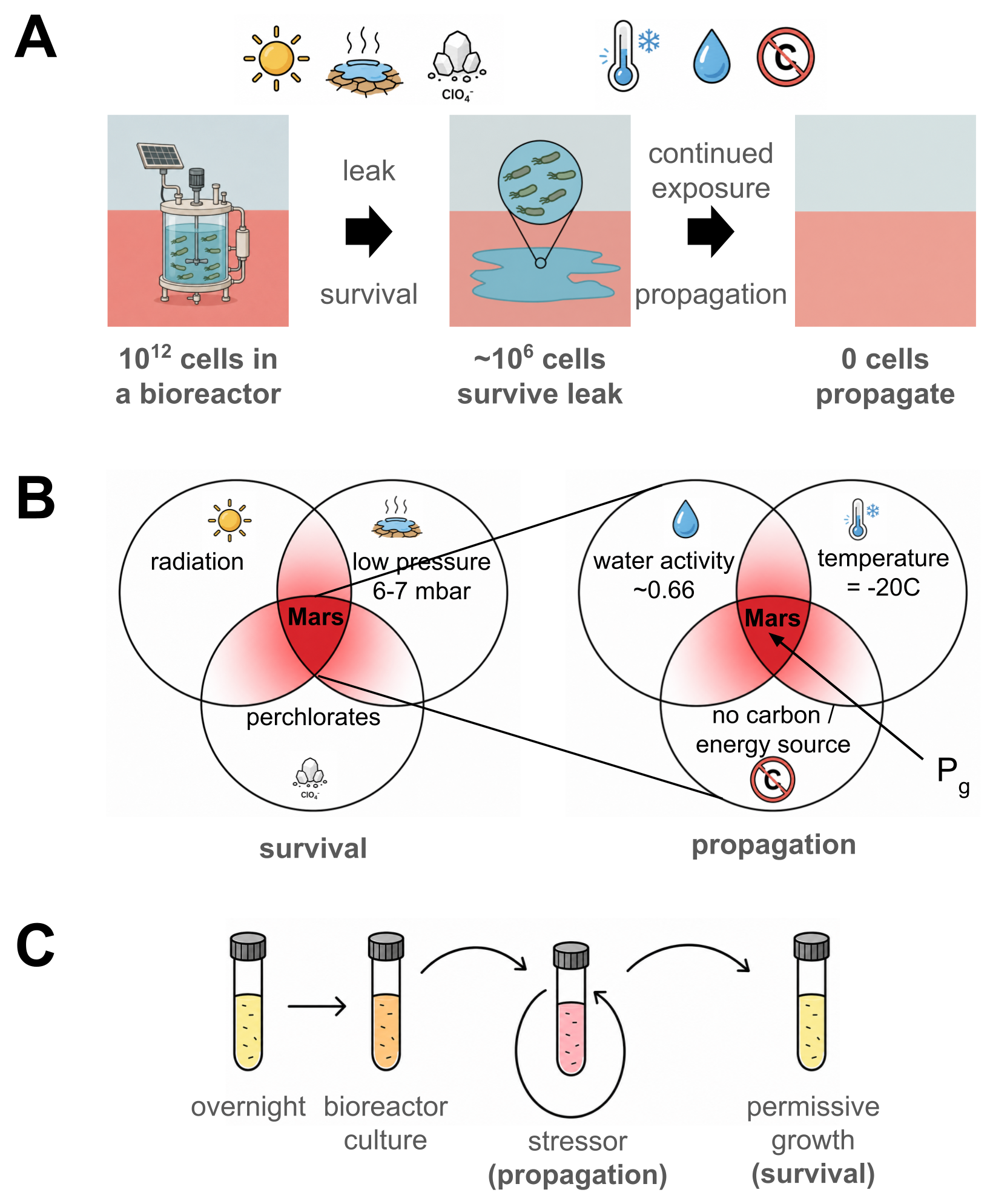}{0.60}{The PRIM framework scenario and assays}{\textbf{A}. A potential forward contamination event, depicted by a high density bioreactor catastrophic leak, is protected by using a PRIM-designated biomanufacturing strain that both experiences a low survival rate and a lack of propagation.\\[2pt] \textbf{B}. The PRIM-designated strain is susceptible to various environmental stressors found on Mars that affect both survival and propagation rates. If these can be experimentally measured, they can produce an estimate of the probability of growth ($P_g$).\\[2pt] \textbf{C}. The PRIM framework assays assess the ability of strains to propagate over the duration of an applied stressor and their survival rate at the end of an applied stressor period.}

\textbf{The strains.} We applied PRIM to two engineered bio-ISRU chassis, sPL.001
and sPL.002 (\textbf{Materials and Methods}). Both are
adaptive-laboratory-evolution isolates that were selected over months of aerobic
passaging\textsuperscript{7} in defined Mars media
(DMM)\textsuperscript{36} with acetate as
the sole carbon source at 30$^{\circ}$C with water activity near 1. Both strains
carry mutations that improve their growth in these simulated Mars conditions.
However, both strains are mesophilic gram-negative heterotrophs that grow optimally
near 30$^{\circ}$C at high water activities; neither sporulates, and neither can
solely use energy sources available at the Martian surface.

\textbf{The assays.} In this initial application of the PRIM framework we measure
$P_g$ for two stressors, carbon starvation and low water activity, in isolation, and
in combination for low water activity at $\mhyphen 20^{\circ}$C. We performed both
propagation and survival assays on these two strains, individually quantifying their
survival and growth under separate Martian stressors (\textbf{Figure 1C}). We
quantified cell densities using optical density (OD), quantitative viability
plating, and most probable number (MPN)
assays\textsuperscript{37} in permissive
media at time points throughout each set of experiments (\textbf{Materials and
Methods}). The propagation assays used a perchlorate-free formulation of DMM to
eliminate perchlorate as a potential stressor. The survival assays used standard,
perchlorate-containing DMM. In both cases, cells were cultivated aerobically and the
DMM concentrations used were permissive of rapid growth.

Propagation assays were conducted for carbon starvation at 30$^{\circ}$C, low water
activity in a chaotropic condition at 30$^{\circ}$C, and low water activity in a
solution with balanced tropicity at -20$^{\circ}$C. In no case was growth observed, as
indicated by increased OD, during the assay, suggesting a lack of propagation in
each condition. Therefore, the reported $P_g$ values are the assay's LOD to 95\%
confidence (\textbf{Table 1}, \textbf{Supplementary Results}). The initial seeded
cell number was counted by a combination of viability plating and OD.

\setlength{\textfloatsep}{18pt plus 4pt minus 2pt}
\setlength{\floatsep}{18pt plus 4pt minus 2pt}
\begin{table*}[b]
\centering
\vspace*{30pt}
\piotablecaption{Propagation assay results.}{The assayed stressor, duration of the experimental condition, and calculated probability of growth ($P_g$) per strain for high-density cultures after prolonged exposure to the stressor. The $P_g$ for $N$ cells is a one-sided 95\% upper confidence bound, or $3/N$. $N$ was calculated by adding together colony forming units (CFU) of replicate cultures exposed to the same strain and condition based on OD or spot plating. The reported $P_g$ accounts for the full volume and density of cells exposed across replicates.}
\begin{tabularx}{\textwidth}{P{0.20\textwidth}P{0.09\textwidth}P{0.16\textwidth}LL}
\piohead{Stressor} & \piohead{Duration} & \piohead{CFU Metric} & \piohead{sPL.001 $P_g$} & \piohead{sPL.002 $P_g$} \\
Carbon starvation & 69 days & OD-based & $\leq 1.2\times10^{\mhyphen 11}$ & $\leq 2.7\times10^{\mhyphen 12}$ \\
\hline
$a_w$ 0.685, chaotropic & 56 days & Spot plating-based & $\leq 1.5\times10^{\mhyphen 13}$ & $\leq 3.3\times10^{\mhyphen 13}$ \\
\hline
$a_w$ 0.667 chaotropically balanced, \mbox{-20°C} & 55 days & OD-based & $\leq 6.3\times10^{\mhyphen 12}$ & $\leq 1.4\times10^{\mhyphen 11}$ \\
\end{tabularx}
\end{table*}

Survival assays were performed under carbon starvation, freeze-thaw, and low water
activity conditions, with a boil condition as a positive control. The survival
fraction is then calculated as the number of cells surviving after exposure to the
stressor divided by the number present before exposure and reported as a unitless
ratio (cells that survive or propagate / cells in the initial population). Survival
was quantifiable for a single freeze-thaw, starvation, and balanced tropicity water
activity (4.67~M glycerol, 1.2~M NaCl, 0.13~M KCl, 0.73~M sucrose),
with a large fraction of cells surviving in each assay. However, boiling and a
chaotropic low-water-activity condition (7.48~M glycerol) were both highly
lethal, with no cells surviving the assay. Therefore, the numbers reported are the
assay's LOD (\textbf{Table 2}, \textbf{Supplementary Results}).

\begin{table*}[b]
\centering
\vspace*{30pt}
\piotablecaption{Survival assays.}{The assayed stressor, duration of the experimental condition, and survival fraction ($N^{surviving}/N^{initial}$) per strain for high-density cultures after brief or prolonged exposure to the stressor.}
\begin{tabularx}{\textwidth}{P{0.16\textwidth}P{0.08\textwidth}P{0.28\textwidth}LL}
\piohead{Stressor} & \piohead{Duration} & \piohead{CFU Metric} & \piohead{sPL.001 Survival Fraction} & \piohead{sPL.002 Survival Fraction} \\
95-98$^{\circ}$C & 20 minutes & MPN LOD & $\leq 10^{\mhyphen 6}$ & $\leq 10^{\mhyphen 6}$ \\
\hline
-80°C & 1 hour & MPN assay & $3.8\times10^{\mhyphen 2}$ & $3.0\times10^{\mhyphen 1}$ \\
\hline
Carbon starvation & 69 days & Numerator: spot plating-based \newline Denominator: OD-based & $5.7\times10^{\mhyphen 1}$ & $1.6\times10^{\mhyphen 1}$ \\
\hline
$a_w$ 0.685, chaotropic & 56 days & No viable cells, reported as 95\% CI & $\leq 1.5\times10^{\mhyphen 13}$ & $\leq 3.3\times10^{\mhyphen 13}$ \\
\hline
$a_w$ 0.667 chaotropically balanced, \mbox{-20°C} & 55 days & Numerator: spot plating-based \newline Denominator: spot plating-based & $8.5\times10^{\mhyphen 3}$ & $3.0\times10^{\mhyphen 4}$ \\
\end{tabularx}
\end{table*}

\begingroup
\setlength{\abovedisplayskip}{10pt plus 2pt minus 1pt}
\setlength{\belowdisplayskip}{10pt plus 2pt minus 1pt}
\setlength{\abovedisplayshortskip}{6pt plus 2pt}
\setlength{\belowdisplayshortskip}{7pt plus 2pt minus 1pt}
\textbf{PRIM calculation.} To calculate the maximum possible $P_g$ consistent with
our data, we chose to use the carbon starvation and chaotropic low-water-activity
condition propagation experiments. We assume these two stressors prevent
propagation in individual ways, as the absence of a reducing source does not share
common mechanisms with the absence of available water. We do not use the
-20°C result, which shares the water-activity stressor. Our calculations
are as follows:
\[
P_c = N_0 \cdot P_g^{\,starvation} \cdot P_g^{\,low\ water\ activity}
\]

\newpage
For sPL.001 our results indicate that
\[
\begin{aligned}
P_g &= P_g^{\,starvation} \cdot P_g^{\,low\ water\ activity} \\
    &= (1.2\times10^{\mhyphen 11})(1.5\times10^{\mhyphen 13}) = 1.8\times10^{\mhyphen 24}
\end{aligned}
\]
Therefore, a maximum number of cells that allows $P_c \leq 10^{\mhyphen 4}$ is
\[
\begin{aligned}
P_c &= N_0 \cdot 1.8\times10^{\mhyphen 24} \leq 10^{\mhyphen 4} \\
N_0 &\leq 5.6\times10^{19}
\end{aligned}
\]
For sPL.002, our results indicate that
\[
\begin{aligned}
P_g &= P_g^{\,starvation} \cdot P_g^{\,low\ water\ activity} \\
    &= (2.7\times10^{\mhyphen 12})(3.3\times10^{\mhyphen 13}) = 9\times10^{\mhyphen 25}
\end{aligned}
\]
Therefore, a maximum number of cells that allows $P_c \leq 10^{\mhyphen 4}$ is
\[
\begin{aligned}
P_c &= N_0 \cdot 9\times10^{\mhyphen 25} \leq 10^{\mhyphen 4} \\
N_0 &\leq 1\times10^{20}
\end{aligned}
\]
At a high-density production target of $10^{11}$ cells/mL, the calculated maximum
allowable sPL.001 $N_0 \leq 5.6\times10^{19}$ corresponds to $5.6\times10^5$~L, or
about 560~m$^3$ of dense culture. A 1,000~L bioreactor, a standard industrial scale,
is more than two orders of magnitude below that limit. This margin is also large
enough to absorb a partial failure of the independence assumption. Including the
interaction term in the above calculation:
\[
P_c = N_0 \cdot \varepsilon \cdot P_g^{\,starvation} \cdot P_g^{\,low\ water\ activity}
\]
The $P_c$ of a 1,000~L bioreactor remains below $10^{\mhyphen 4}$ up to interaction terms
of $\varepsilon \approx 560$, well above microbial growth interaction terms reported
in the literature\textsuperscript{26-28}.
\endgroup

\newpage
\section{Extensions of the PRIM Framework}

Above, we shared experimental data for two Mars bio-ISRU chassis strains that could
be intentionally introduced to the Martian environment for high biomass generation
and bioproduction. We showed that even in a catastrophic high biomass leak
situation, these two strains could be given PRIM designation as $P_c$ remains
$\leq 10^{\mhyphen 4}$. Future development of the PRIM framework could extend beyond this
use case. Here, we present considerations for 1) how non-worst case scenarios could
be calculated and 2) how the PRIM framework could be extended to other
organisms-of-interest for intentional use on Mars.

\textbf{Non-worst case scenarios}

Above we have only considered a couple environmental barriers to growth on Mars; in
reality, many additional safeguards would be in place to prevent the possibility of
a spill event through careful engineering of the bioreactor device and other
measures. We simplified the Coleman-Sagan equation to just the number of organisms
($N_0$) and the probability of their growth ($P_g$) by imagining a scenario where
all other variables $=1$. In reality, additional mitigation strategies would be in
effect to reduce the probability of release ($P_R$) below 1, while reduction ($R$),
probability of an organism reaching the surface ($P_S$), and probability of impact
($P_I$), or leakage in this case, would remain equal to 1. The probability of
release could be mitigated through secondary and tertiary containment strategies,
including physical separation from the environment, the release of additional
biocidal agents, and engineered microbial kill switches to further reduce $P_c$ if
necessary.

\subsection{Other organisms}

In our worked example, we presented experimental data to assess $P_g$ for known
clonal isolates. However, experimental measurement of the upper bound on $P_g$ may
not be feasible for all organisms. For example, the human gut or skin microbiome is
not a well-defined community with a fixed number of species, and assumptions would
need to be made about what representative organisms to include and how to culture
them appropriately. Similar characterization of entire organisms like plants may be
infeasible due to the long lifecycle of plants. For organisms beyond heterotrophs,
which require organic carbon, cannot photosynthesize, and do not sporulate, we
recommend tuning of the PRIM framework to prove that the probability of growth
($P_g$) is sufficiently low. Below, we outline future considerations for extending
the PRIM framework to these cases:

\begin{itemize}[leftmargin=1.2em, itemsep=0.08em, topsep=0.1em, parsep=0pt, partopsep=0pt, before=\setlength{\parskip}{0pt}]
  \item \textbf{Phototrophs} cannot be starved on the Martian surface because
    they fix their own carbon from CO\textsubscript{2} and draw energy from
    sunlight. In lieu of starvation assays, UVB/UVC irradiation, both at the
    surface and under a light dust cover that still allows visible light, could
    be used, as these organisms require light exposure to engage in
    photosynthesis. It is possible that certain minerals could attenuate
    damaging UV radiation while transmitting enough photosynthetically active
    radiation for microbial growth\textsuperscript{38,39}.
  \item \textbf{Anaerobes} would require testing conditions in anoxic
    headspaces to prove that their energy metabolism is not compatible with
    the combined Martian atmosphere and regolith. While low water activity
    should still be extremely biocidal, anaerobes will likely be more
    difficult to assay to a low $P_g$.
  \item \textbf{Biofilms} are widely regarded as protective aggregate
    structures\textsuperscript{40}.
    Biofilm-forming organisms might be able to survive to higher cell counts
    than their planktonic counterparts, especially against stressors like
    desiccation and UV irradiation\textsuperscript{41}.
    Either the ability to form biofilms can be knocked out of potential Mars
    bio-ISRU organisms or the PRIM framework needs to be performed on both
    planktonic and biofilm cultures.
  \item \textbf{Spore forming organisms} are called out directly in planetary
    protection policy. Bacterial endospores survive desiccation, UV, oxidative
    stress, and temperature extremes far beyond the vegetative cell, and are
    the reference hazard against which spacecraft bioburden is
    measured\textsuperscript{11,42,43}.
    Though the probability of survival of spores is higher, their probability
    of propagation is likely still low as germination and subsequent outgrowth
    often requires particular chemical signals, sufficient hydration, and other
    permissive environmental factors\textsuperscript{44}.
    All together, spore-formers are more complicated organisms for assaying
    $P_g$ and other metrics, such as quantifying spore germination and
    outgrowth, should be considered.
  \item \textbf{Extremophiles} overall are less likely to achieve PRIM
    designation because of their robustness to specific environmental
    stressors. They can be tested in similar conditions, but might not be good
    chassis for bio-ISRU without more specific containment mechanisms. In
    particular, aeonophiles\textsuperscript{45},
    a new class of extremophiles whose metabolism and cell division occurs
    over long periods of time, would be hard to assay in these benchtop and
    time limited experiments.
  \item \textbf{Microbial communities} should be assayed directly, as if the
    community were a single clonal strain, rather than inferred from its
    members tested in isolation. A community can be resilient in ways via
    metabolic cross-feeding or aggregate structure (\textit{i.e.}\ the biofilm argument
    applied to a consortium). Running the survival and propagation assays on
    the intact community bounds the community-level $P_g$. The assay bounds
    the community as tested, not every sub-population it might later resolve
    into. Individual strains that might be highly resilient to Martian
    stressors should be assayed as clonal isolates if they are able to grow
    independently.
  \item \textbf{The subsurface of Mars} could potentially house habitable
    conditions; these have been neither proven nor
    disproven\textsuperscript{46,47}.
    More work could be done to estimate probabilities of their existence in
    addition to the probability of leakage of cell culture down to these
    sites. Bio-ISRU strains should not be chemolithoautotrophs until more is
    known about the subsurface chemistry.
\end{itemize}

\section{Implications for Planetary Protection}

The goal of ongoing conversations within the astrobiology and space exploration
communities is to adapt planetary protection frameworks to enable human exploration
of Mars while maintaining protective forward and backward contamination measures.
For forward contamination in particular, the current COSPAR planetary protection
policies minimize input bioburden because, for an unknown contaminant, $P_g$ cannot
be predicted. The identity and physiology of clean room contaminants are not fully
characterized, so the variable that can be reduced is how many of them ($N_0$)
hitchhike to Mars. Here, we described the PRIM designation for a strain that has
cleared the survival and propagation assays outlined above, and is thereby shown
unable to propagate on the Martian surface to a one-in-ten-thousand margin
($P_c \leq 10^{\mhyphen 4}$), even under a worst-case high-titer release in which the full
bioreactor contents leak and every other term in the Coleman-Sagan formula is set to
its worst-case value. PRIM does not replace bioburden control; it adds a second,
measurement-based path to quantifiable compliance for the specific case where the
organism is known. This matters now because the next wave of missions will send
biology intentionally. Bio-ISRU bioreactors are a near-term case that could be
demonstrated before human missions, but the throughline runs to crewed missions,
which will carry biological life support systems (BLSS) and human-associated
microbial communities, both of which cannot be sterilized.

PRIM is currently a policy proposal focused on mitigating propagation-related
forward contamination of bio-ISRU organisms. The PRIM framework is a first step
towards developing policy to quantify and estimate chassis-specific, and eventually
community-specific, risk. The assays, the limit-of-detection margins, and the
choices of the most permissive Mars conditions all need scrutiny from the planetary
protection community before it can be accepted as a community standard. We welcome
all engagement and feedback with this proposal.

\section{Materials and Methods}

\subsection{Strains}

We obtained the following bacterial strains from Leibniz Institute DSMZ-German
Collection of Microorganisms and Cell Cultures (DSMZ): \textit{Cupriavidus necator}
H16 (DSM 428) and \textit{Pseudomonas putida} KT2440 (DSM 6125). We grew them up in
their recommended media and glycerol stocked them for long term -80°C
storage.

Both of these strains were subsequently engineered for Mars bio-ISRU relevance\textsuperscript{7}. Briefly, the \textit{C. necator} strain underwent a
functional genomics screen that added a 3.4~kb genomic insertion from the
\textit{Deinococcus radiodurans} R1 (American Type Culture Co\-llection, ATCC 13939)
genome, consisting of one uncharacterized gene and a kanamycin resistance marker.
50~\textmu g/mL kanamycin was added to all cultures derived from this strain.
Afterwards, both this engineered \textit{C. necator} strain and wild-type
\textit{P. putida} underwent adaptive laboratory evolution (ALE) for improved growth
in Mars bio-ISRU conditions and we refer to the resulting strains as sPL.001 and
sPL.002, respectively. For all of these experiments, the strains were grown in a
defined Mars media (DMM) and Mars trace micronutrients (MTM) with potassium acetate
added as a sole carbon source\textsuperscript{36}.
The concentrations of DMM and potassium acetate were increased to raise the
selective pressure that the strains were exposed to. Whole-genome sequencing of
isolates post-ALE identified mutations in pathways related to cell division,
metabolism, and growth. These divergences from the wild-type strains improved growth
in DMM.

\subsection{Growth conditions and condition variability}

To test these strains, we grew 5~mL overnight cultures in Super Optimal Broth (SOB)
or Luria Broth (LB). For each propagation and starvation assay described below,
overnight cultures were washed two to three times in 10~mM TES pH 8 + 1\% NaCl,
referred to as TES-buffered saline (TBS), and used to inoculate a preculture in
DMM + MTM + TES + potassium acetate. In precultures, strains were grown in DMM at
different concentrations for the following propagation and survival assays. Media
are designated $n\times$ DMM, where $n$ is the multiple of the 1X DMM macronutrient
and micronutrient concentrations, followed by additives. Differences in growth
conditions (media, volumes, stressors) are shown in detail in \textbf{Supplementary
Table 1}.

\subsection{Starvation propagation and survival assays}

Starvation (\textit{i.e.}\ no added carbon) propagation and survival assays were measured in
a single continuous 69-day experiment in which the same cultures provided both the
propagation time course and the terminal survival endpoints.

For each strain, overnights grown in SOB were used to inoculate 2~L of preculture to
an OD of $\sim 0.013$. These precultures were grown overnight at 30$^{\circ}$C
with shaking. 1.5~L per strain was pooled, harvested by centrifugation, washed three
times with TBS, and divided into three equal portions for seeding into the
downstream assays. Three 500~mL cultures were prepared and seeded with one portion
of washed preculture. Cultures were held at 30$^{\circ}$C with shaking for 69 days
with no medium replenishment or feeding. ODs measured immediately after seeding were
OD 0.226-0.280 (sPL.001) and 1.131-1.151 (sPL.002), corresponding at an assumed
$10^9$ cells~mL$^{\mhyphen 1}$ OD$^{\mhyphen 1}$ to 1.1-1.4$\times10^{11}$ and 5.7$\times10^{11}$
cells per culture, or 2.5$\times10^{11}$ and 1.1$\times10^{12}$ cells when pooled
across the two starvation replicates of each strain.

Population density was measured weekly by OD through day 26 and by spot plating
through day 69. For spot plating counts, 1~mL of media was withdrawn and serially
diluted before 5~\textmu L was spotted onto SOB agar trays. Colonies were counted
after one and two days of growth at 30$^{\circ}$C. The most concentrated spot was
undiluted culture, so a single colony at that position corresponds to 200~colony
forming units (CFU) mL$^{\mhyphen 1}$. Day 0 cell numbers are derived from OD
(1~OD = $10^9$ cells mL$^{\mhyphen 1}$, uncalibrated) because spot-plating counts
from those samples were rejected due to a dilution series error during the survival
assay. Survival fractions are the day 69 viable count
divided by the day 0 viable count for the same culture; the higher survival of the
two replicates is reported.

\subsection{Freeze-thaw / boil survival assay}

Overnight SOB cultures were washed and normalized to OD 0.7, and used to inoculate
3~mL of preculture. After 23 hours at 30$^{\circ}$C with shaking, cultures were
re-normalized to OD 0.7 in the same medium ($\sim 8\times10^8$ cells~mL$^{\mhyphen 1}$) and
split into three 1~mL aliquots in 2~mL tubes: unstressed (room temperature, 2~h), a
single freeze-thaw cycle (-80°C for 1~h, then 1~h thaw at room
temperature), and boiled (95-98$^{\circ}$C, 20~min). The boil condition served as
a positive control for reduced cell viability.

Viable cells were enumerated by a most probable number (MPN) assay, whereby the
experimental samples were serially diluted eight times, with four replicate wells per
dilution, inoculated into SOB medium, then incubated three days at 30$^{\circ}$C with
shaking and scored as dense growth or no growth. MPN was estimated by maximum
likelihood under a Poisson single-hit model with 95\% confidence intervals from the
profile-likelihood ratio; the detection floor for this plate layout was 40 cells
mL$^{\mhyphen 1}$ in experimental culture. Survival is expressed as
MPN(stressed)/MPN(unstressed). The same dilution series was also spot plated to
validate results.

\subsection{Low temperature propagation and survival assays}

A liquid medium is required to test propagation at -20°C, and a depressed
freezing point also means depressed water activity. The cold assay is therefore a
combined cold plus low-water-activity condition, and is excluded from the
independent product in the PRIM calculation. However, a solution with balanced
tropicity was used to remove chaotropism as a stressor in this condition.

Overnights in SOB were washed once in TBS and used to seed 2~L per strain of
preculture grown at 30$^{\circ}$C overnight. These precultures were harvested by
centrifugation, washed once in TBS, and separated into three portions.

Each 375~mL replicate was seeded with one portion of washed preculture, with one
room temperature positive control and two independent cold-temperature replicates
per strain. The medium used for the cold-temperature was brought to $a_w$ 0.667
with 4.67~M glycerol, 1.20~M NaCl, 0.13~M KCl and 0.73~M sucrose, a solute mixture
selected to be chaotropically balanced at that water
activity\textsuperscript{48}. The
positive control was 5X DMM + 25~mM TES pH 8 + 1\% w/v potassium acetate, held at
room temperature of $\sim 20^{\circ}$C for the duration of the experiment without
shaking. No deliberate thermal cycling was applied; vessels were removed from
-20°C briefly for sampling.

Population density was followed by optical density (OD) and by spot plating as
described above through day 55. Spot plating was performed as described above.
Colonies were counted after one and two days of growth at 30$^{\circ}$C. Viable
count rather than OD is the reported metric for this experiment. Survival fractions
are the terminal viable count divided by the day-0 viable count for the same
culture; the higher survival of the two replicates is reported.

\subsection{Low water activity propagation and survival assays}

Water activity propagation and survival were measured in a single continuous 56-day
experiment, in which the same cultures provided both the propagation time course
and the terminal survival endpoint. The solute system used was 7.48~M glycerol, for
an $a_w$ of 0.685 with high chaotropicity mimicking the environment of Martian
brines\textsuperscript{48}. The
additional DMM salts, acetate, and TES present in these media are expected to depress
$a_w$ by less than 0.01 relative to the cited values, and is not an important factor.

Precultures were conducted identically to the cold propagation assay. SOB
overnights were washed once in TBS and used to seed 2~L per strain grown at
30$^{\circ}$C overnight. These cultures were harvested by centrifugation, washed
once in TBS, and separated into three portions. Each 375~mL replicate was seeded
with one portion of washed preculture, with one normal culture positive control and
two independent low $a_w$ replicates per strain. The positive control was the same
medium without glycerol. Cells exposed in the low-$a_w$ conditions were determined by
viability plating rather than OD, since OD was found to be unreliable with the
glycerol additive.

Viable cell counting was conducted by spot dilution through day 56. However, the
number of viable cells fell below the detection limit of the spot plating method,
which is $10^5$ cells in each culture. The spot plating was continued in case any
cells were able to replicate and rise above the detection limit again. At the end
of the assay, the entire culture was recovered and plated. Each culture was
centrifuged in full (4,000$\times g$, 45~min), the pellet resuspended in
$\sim$10~mL of the same medium, re-centrifuged, resuspended in 2~mL TBS, and
bead-spread onto large LB Q-trays without antibiotic, incubated at 30$^{\circ}$C
and read at 1, 2 and 5 days. Recovered colonies were restreaked and identified by
16S sequencing to confirm identity. The viscosity of the glycerol media caused
occasional failed transfers during automated spot plating; affected wells were
identified as blank rows against parallel rows and excluded from analysis. The
final whole-culture plating was not subject to this failure mode.

\subsection{PRIM calculations}

\textbf{Cells exposed.} The number of cells exposed in each culture, $N_0$, was
determined by viability plating or OD measurement. Where day-0 plate data were
reliable, viability plating was used as described above. Where contamination or
pipetting errors occurred, OD at an assumed $10^9$ cells mL$^{\mhyphen 1}$ OD$^{\mhyphen 1}$ was
used. OD was used for the starvation and cold conditions and viability plating for
the chaotropic low-water-activity condition. For the propagation assays, $N_0$ was
summed across replicate cultures of the same strain and condition.

\textbf{Limits of detection.} According to the statistical rule of three, for zero
observed events in a Poisson process, the 95\% confidence limit for an upper bound
is given as $3/N$, where $N$ is the number of trials. Therefore, our estimate of the
lower LOD for the propagation assays is given as $3/N_0$. All limits of detection
reported here are computed on that basis. A LOD of $3/N_0$, or one survivor anywhere
in the exposed population, was established by one of two routes. For the chaotropic
low-water-activity condition, it was determined by terminal whole-culture recovery,
which interrogates the entire exposed volume for any trace of survival. For the
starvation and cold conditions, continuous observation showed no increase in cell
density. Viable cell counts remained above the spot-plating assay floor at every
timepoint and only decreased. If any one of the $N_0$ cells present at the beginning
of the assay were able to sustain propagation, it would have produced a net increase
in that count over the two-month duration of the assay, and none was observed.

\textbf{Product of independent propagation limits.} Carbon starvation and
chaotropic low-water-activity were treated as independent stressors and their
propagation limits were therefore multiplied. The $a_w = 0.667$ and
-20°C condition is excluded from the product because it is a combination
of both low-water-activity and chaotropicity, and is therefore not independent of
it.

\section{Reviewers}

The Pioneer Labs team thanks the following external reviewers for lending their
expertise in planetary science, microbiology, and astrobiology. Their role was to
give candid feedback throughout the study, helping us sharpen our rationale and
experimental design. Their input shaped our methods and/or writing, but they were
not asked to endorse our conclusions and did not review the final draft.

\begin{itemize}[leftmargin=1.2em, itemsep=0.2em, topsep=0.2em]
  \item Alfonso Davila (NASA Ames Research Center)
  \item Mario Maggio (Ambrosia Space)
  \item Cyprien Verseux (Center of Applied Space Technology and Microgravity
    (ZARM), University of Bremen)
  \item Erika Wagner (The Exploration Company)
  \item Maria-Paz Zorzano Mier (Centro de Astrobiolog\'ia (CSIC-INTA))
\end{itemize}

\section{Acknowledgements}

This project was supported in part by a grant from The Astera Institute. In
compliance with Astera's Open Science
policy\textsuperscript{49}, this paper
will not be submitted to a journal and is presented here in its final form. We thank
Olesia Bushkova for her assistance with scientific communication.

\section{References}

\begin{pioneerrefsnum}
\item National Academies of Sciences, Engineering, Medicine \& Others. \textit{Report Series: Committee on Planetary Protection: Evaluation of Bioburden Requirements for Mars Missions}. (nap.nationalacademies.org, 2021).
\item Fischer, J. \textit{et al.} Lessons Learned from the Integration of Biological Systems in Series for Wastewater Treatment on Early Planetary Bases. in (51st International Conference on Environmental Systems, 2022).
\item Nangle, S. N. \textit{et al.} The case for biotech on Mars. \textit{Nat. Biotechnol.} \textbf{38}, 401-407 (2020).
\item Berliner, A. J. \textit{et al.} Towards a Biomanufactory on Mars. \textit{Front. Astron. Space Sci.} \textbf{8}, (2021).
\item Averesch, N. J. H. \textit{et al.} Microbial biomanufacturing for space-exploration-what to take and when to make. \textit{Nat. Commun.} \textbf{14}, 2311 (2023).
\item Wordsworth, R. \textit{et al.} Applied astrobiology: An integrated approach to the future of life in space. \textit{Astrobiology} \textbf{25}, 327-330 (2025).
\item Martin, F. R. \textit{et al.} Bioplastic Production from simulated 100\% in situ Mars resources. \textit{bioRxiv} (2026) doi:~\href{https://doi.org/10.64898/2026.09.14.748363}{\underline{10.64898/2026.09.14.748363}}.
\item Llorente, B., Williams, T. C., Goold, H. D., Pretorius, I. S. \& Paulsen, I. T. Harnessing bioengineered microbes as a versatile platform for space nutrition. \textit{Nat. Commun.} \textbf{13}, 6177 (2022).
\item United Nations Office for Outer Space Affairs. Treaty on principles governing the activities of states in the exploration and use of outer space, including the moon and other celestial bodies. in \textit{International Space Law} 3-9 (UN, 2018).
\item COSPAR Panel on the Panel on Planetary Quarantine and the Planetary Protection. COSPAR policy on planetary protection. \textit{Space Res. Today} \textbf{211}, 12-25 (2021).
\item National Aeronautics and Space Administration. NASA Planetary Protection Handbook. \textit{NASA Planetary Protection Handbook} \href{https://ntrs.nasa.gov/citations/20240016475}{\underline{https://ntrs.nasa.gov/citations/20240016475}} (2010).
\item Conley, C. A. Coleman-Sagan equation. in \textit{Encyclopedia of Astrobiology} 613-613 (Springer Berlin Heidelberg, Berlin, Heidelberg, 2023).
\item Brener, D. J. \& Cockell, C. S. Rethinking planetary protection: an island biogeographical analysis. \textit{J. R. Soc. Interface} \textbf{22}, 20250079 (2025).
\item Cockell, C. S. Fifty years after Viking: The promise of solar system microbiology and microbial ecology. \textit{Astrobiology} \textbf{26}, 177S-187S (2026).
\item DeVincenzi, D. L., Stabekis, P. D. \& Barengoltz, J. B. A proposed new policy for planetary protection. \textit{Advances in Space Research} \textbf{3}, 13-21 (1983).
\item DeVincenzi, D. L. \& Stabekis, P. D. Revised planetary protection policy for solar system exploration. \textit{Advances in Space Research} \textbf{4}, 291-295 (1984).
\item MEPAG Special Regions-Science Analysis Group. Findings of the Mars special regions science analysis group. \textit{Astrobiology} \textbf{6}, 677-732 (2006).
\item Rummel, J. D. \textit{et al.} A new analysis of Mars 'Special Regions': findings of the second MEPAG Special Regions Science Analysis Group (SR-SAG2). \textit{Astrobiology} \textbf{14}, 887-968 (2014).
\item Olsson-Francis, K. \textit{et al.} The COSPAR Planetary Protection Policy for robotic missions to Mars: A review of current scientific knowledge and future perspectives. \textit{Life Sci. Space Res. (Amst.)} \textbf{36}, 27-35 (2023).
\item Spry, J. A. \textit{et al.} Planetary protection knowledge gap closure enabling crewed missions to Mars. \textit{Astrobiology} \textbf{24}, 230-274 (2024).
\item Committee on a Science Strategy for the Human Exploration of Mars, Space Studies Board, Division on Engineering and Physical Sciences \& National Academies of Sciences, Engineering, and Medicine. A science strategy for the human exploration of mars. Preprint at https://doi.org/\href{http://dx.doi.org/10.17226/28594}{\underline{10.17226/28594}} (2026).
\item Reimagining Planetary Protection for Mars Exploration II Oral at AbSciCon 2026.
\item Limits of Terrestrial Life and Probability of Growth on Mars Virtual Workshop 2026.
\item Kurtz, D. A. \textit{et al.} Real-world limitations to detection. in \textit{ACS Symposium Series} 288-316 (American Chemical Society, Washington, DC, 1987).
\item Hanley, J. A. \& Lippman-Hand, A. If nothing goes wrong, is everything all right? Interpreting zero numerators. \textit{JAMA} \textbf{249}, 1743-1745 (1983).
\item Lambert, R. J. W. \& Bidlas, E. An investigation of the Gamma hypothesis: a predictive modelling study of the effect of combined inhibitors (salt, pH and weak acids) on the growth of Aeromonas hydrophila. \textit{Int. J. Food Microbiol.} \textbf{115}, 12-28 (2007).
\item Lambert, R. J. W. \& Bidlas, E. A study of the Gamma hypothesis: predictive modelling of the growth and inhibition of Enterobacter sakazakii. \textit{Int. J. Food Microbiol.} \textbf{115}, 204-213 (2007).
\item Le Marc, Y. \textit{et al.} Modelling the growth kinetics of Listeria as a function of temperature, pH and organic acid concentration. \textit{Int. J. Food Microbiol.} \textbf{73}, 219-237 (2002).
\item Wadsworth, J. \& Cockell, C. S. Perchlorates on Mars enhance the bacteriocidal effects of UV light. \textit{Sci. Rep.} \textbf{7}, 4662 (2017).
\item Harrison, J. P., Gheeraert, N., Tsigelnitskiy, D. \& Cockell, C. S. The limits for life under multiple extremes. \textit{Trends Microbiol.} \textbf{21}, 204-212 (2013).
\item Pitt, J. I. Xerophilic fungi and the spoilage of foods of plant origin. in \textit{Water Relations of Foods} 273-307 (Elsevier, 1975).
\item Williams, J. P. \& Hallsworth, J. E. Limits of life in hostile environments: no barriers to biosphere function? \textit{Environ. Microbiol.} \textbf{11}, 3292-3308 (2009).
\item Raghavendra, J. B., Zorzano, M.-P. \& Martin-Torres, J. Growth of microorganisms in a Martian regolith simulant at reduced water activity. \textit{Sci. Rep.} \textbf{16}, (2026).
\item Boxe, C. S. \textit{et al.} Adsorbed water and thin liquid films on Mars. \textit{Int. J. Astrobiology} \textbf{11}, 169-175 (2012).
\item Chevrier, V. F. \& Slank, R. A. The elusive nature of Martian liquid brines. \textit{Proc. Natl. Acad. Sci. U. S. A.} \textbf{121}, e2321067121 (2024).
\item Greene, H. \textit{et al.} Defined Mars Media (DMM), a chemically defined simulant of the soluble macro- and micro- nutrients in Mars regolith for use in biological research. \textit{bioRxiv} (2026) doi:\href{http://dx.doi.org/10.64898/2026.04.24.719001}{\underline{10.64898/2026.04.24.719001}}.
\item Martini, K. M., Boddu, S. S., Nemenman, I. \& Vega, N. M. Maximum likelihood estimators for colony-forming units. \textit{Microbiol. Spectr.} \textbf{12}, e0394623 (2024).
\item Wierzchos, J. \textit{et al.} Adaptation strategies of endolithic chlorophototrophs to survive the hyperarid and extreme solar radiation environment of the Atacama Desert. \textit{Front. Microbiol.} \textbf{6}, 934 (2015).
\item Kugler, A. \& Dong, H. Phyllosilicates as protective habitats of filamentous cyanobacteria Leptolyngbya against ultraviolet radiation. \textit{PLoS One} \textbf{14}, e0219616 (2019).
\item Stewart, P. S. \& Franklin, M. J. Physiological heterogeneity in biofilms. \textit{Nat. Rev. Microbiol.} \textbf{6}, 199-210 (2008).
\item Frösler, J., Panitz, C., Wingender, J., Flemming, H.-C. \& Rettberg, P. Survival of Deinococcus geothermalis in biofilms under desiccation and simulated space and martian conditions. \textit{Astrobiology} \textbf{17}, 431-447 (2017).
\item Nicholson, W. L., Munakata, N., Horneck, G., Melosh, H. J. \& Setlow, P. Resistance of Bacillus endospores to extreme terrestrial and extraterrestrial environments. \textit{Microbiol. Mol. Biol. Rev.} \textbf{64}, 548-572 (2000).
\item Horneck, G. \textit{et al.} Resistance of bacterial endospores to outer space for planetary protection purposes\-experiment PROTECT of the EXPOSE-E mission. \textit{Astrobiology} \textbf{12}, 445-456 (2012).
\item Setlow, P., Wang, S. \& Li, Y.-Q. Germination of spores of the orders Bacillales and Clostridiales. \textit{Annu. Rev. Microbiol.} \textbf{71}, 459-477 (2017).
\item Lloyd, K. G. \& Steen, A. D. Defining ultra-slow-growing extremophilic microorganisms as aeonophiles. \textit{Nat. Microbiol.} \textbf{10}, 1555-1557 (2025).
\item Michalski, J. R. \textit{et al.} Groundwater activity on Mars and implications for a deep biosphere. \textit{Nat. Geosci.} \textbf{6}, 133-138 (2013).
\item Tarnas, J. D. \textit{et al.} Earth-like habitable environments in the subsurface of Mars. \textit{Astrobiology} \textbf{21}, 741-756 (2021).
\item Stevenson, A. \textit{et al.} Glycerol enhances fungal germination at the water-activity limit for life. \textit{Environ. Microbiol.} \textbf{19}, 947-967 (2017).
\item Astera Institute. Astera Open Science Policy. Preprint at https://doi.org/\href{http://dx.doi.org/10.5281/ZENODO.17873235}{\underline{10.5281/ZENODO.17873235}} (2025).
\end{pioneerrefsnum}

\clearpage
\onecolumn
\section{Supplementary Results}

Full descriptions of the stressors, media, volumes, and duration of the various
propagation and survival assays (\textbf{Materials and Methods}) are described in
full in \textbf{Supplementary Table 1}.

\par\vspace{6pt}
\begin{center}
\piotablecaptionnonum{Supplementary Table 1.}{A full summary of the tested stressors, the media and volumes used for different aspects of their associated assays, the duration of the assays themselves, and information on the reported metrics for $P_g$. In the table, DMM = defined Mars media, MTM = Mars trace micronutrients, and KOAc = potassium acetate.}
\begin{tabularx}{\textwidth}{P{0.11\textwidth}LLLL}
\piohead{Parameter} & \piohead{Starvation} & \piohead{Freeze-thaw/boil} & \piohead{Low temperature (-20°C)} & \piohead{Low water activity ($a_w$)} \\
\textbf{Description} & No added carbon source & Single freeze-thaw cycle at -80°C for 1 h and room temperature at 1 h; boil (95-98$^{\circ}$C) 20 min & Glycerol (low $a_w$) + growth at -20°C & Low $a_w \sim 0.685$ with high chaotropicity \\
\hline
\textbf{Preculture medium} & 10X DMM + 10X MTM + 1\% potassium acetate (KOAc) + 50~mM TES pH 8 & 10X DMM + 10X MTM + 1\% KOAc + 50~mM TES pH 8 & 5X DMM + 5X MTM + 1\% KOAc + 25~mM TES pH 8 & 5X DMM + 5X MTM + 1\% KOAc + 25~mM TES pH 8 \\
\hline
\textbf{Perchlorate concentration} & None & 10X DMM levels ($\sim$30~mM) & None & None \\
\hline
\textbf{Assay medium} & Same as preculture without KOAc & Same as preculture, with 1\% (sPL.001) or 1.5\% (sPL.002) KOAc & Same as preculture, with 4.67~M glycerol, 1.20~M NaCl, 0.13~M KCl and 0.73~M sucrose to $a_w \sim 0.667$ & Same as preculture, with 7.48~M glycerol, $a_w \sim 0.685$ \\
\hline
\textbf{Positive control} & Same as assay medium with 1\% KOAc added & No exposure to freeze-thaw or boil & Same as assay medium without glycerol at 20$^{\circ}$C & Same as assay medium without glycerol \\
\hline
\textbf{Volume per replicate} & 500~mL & 1~mL & 375~mL for replicates; 250~mL for positive control & 375~mL for replicates; 250~mL for positive control \\
\hline
\textbf{Replicate design} & 2 replicates, 1 positive control & 1 replicate, 1 positive control & 2 replicates, 1 positive control & 2 replicates, 1 positive control \\
\hline
\textbf{Duration} & 69 days & 2 h stress window & 55 days & 56 days \\
\hline
\textbf{Enumeration} & OD to day 26; spot plating to day 69 & MPN assay + spot plating to validate & Spot plating to day 55 & Spot plating, then whole-culture recovery \\
\hline
\textbf{Reported metric} & Viable count & MPN$_{stressed}$ / MPN$_{positive\ control}$ & Viable count & Viable count \\
\end{tabularx}
\end{center}
\par\vspace{8pt}

\clearpage
\subsection{Starvation propagation assays}

The carbon source added (+acetate) positive control grew from OD 0.275 to 1.825
(sPL.001) and 1.151 to 1.65 (sPL.002) in the first 24 hours, confirming that the
base medium supported growth from the same inoculum, then the OD declined to the
assay floor as acetate was exhausted. The OD of the starvation samples declined from
day 0 with no growth phase in either strain. Viable cell counts remained above the
spot-assay floor at every timepoint, so the populations were under continuous
observation for the full exposure period.

\piofigurelabeledinline{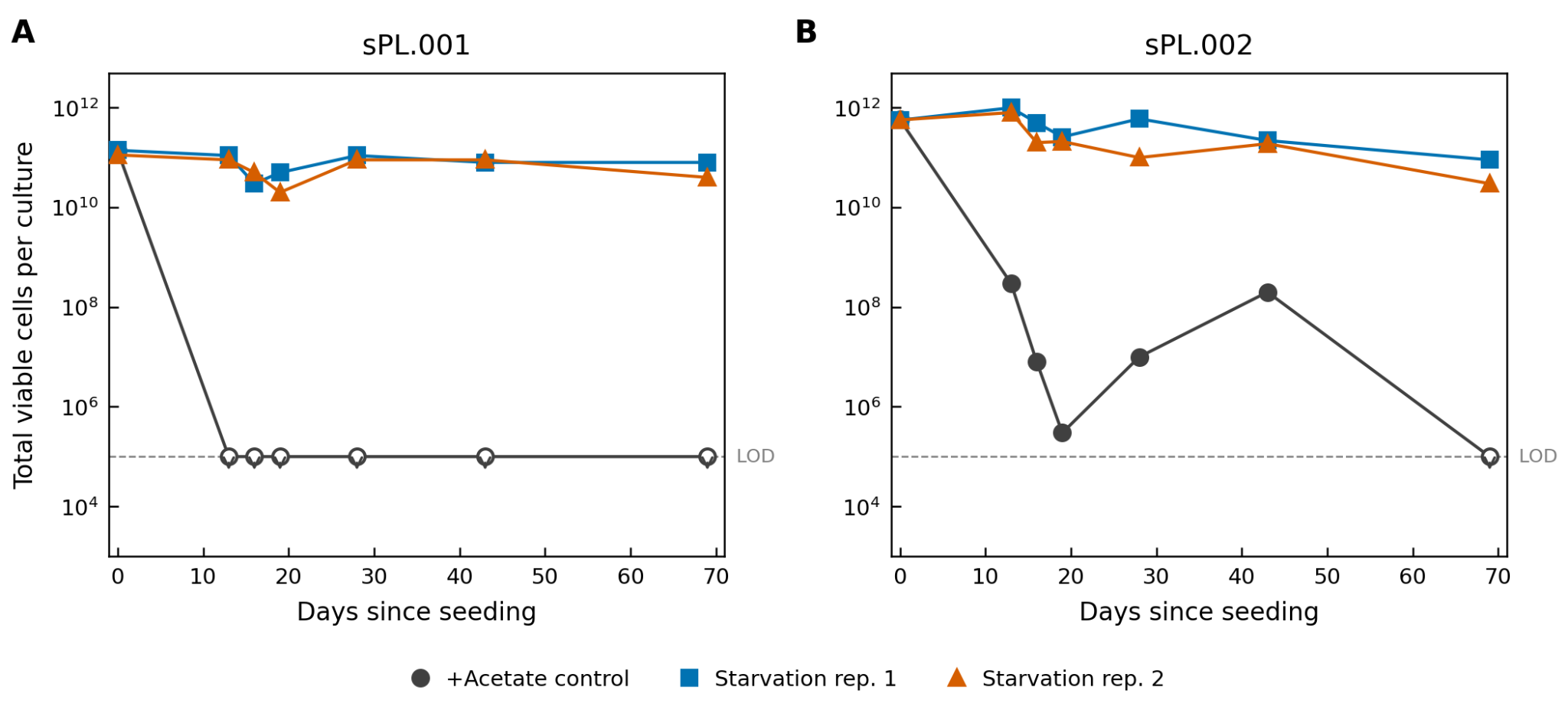}{S1}{Starvation propagation by spot plating or optical density.}{Total viable cells per culture (log\textsubscript{10} scale) over 69 days for sPL.001 (\textbf{A}) and sPL.002 (\textbf{B}). Two replicates (blue squares and orange triangles) and one positive control (black circles) are shown. The horizontal dashed line indicates the assay LOD (one colony in a 5~\textmu L spot = $1\times10^5$ total CFU per 500~mL culture). Open markers with downward arrows denote samples plotted at the LOD. Day 0 total viable cells were calculated using an OD-to-CFU conversion; the rest of the values are based on spot plating colony counting. The sPL.001 positive control reached the LOD by day 13 as acetate was exhausted. Neither starvation condition of either strain showed sustained growth over the 69-day observation period.}

\piofigurelabeledinline{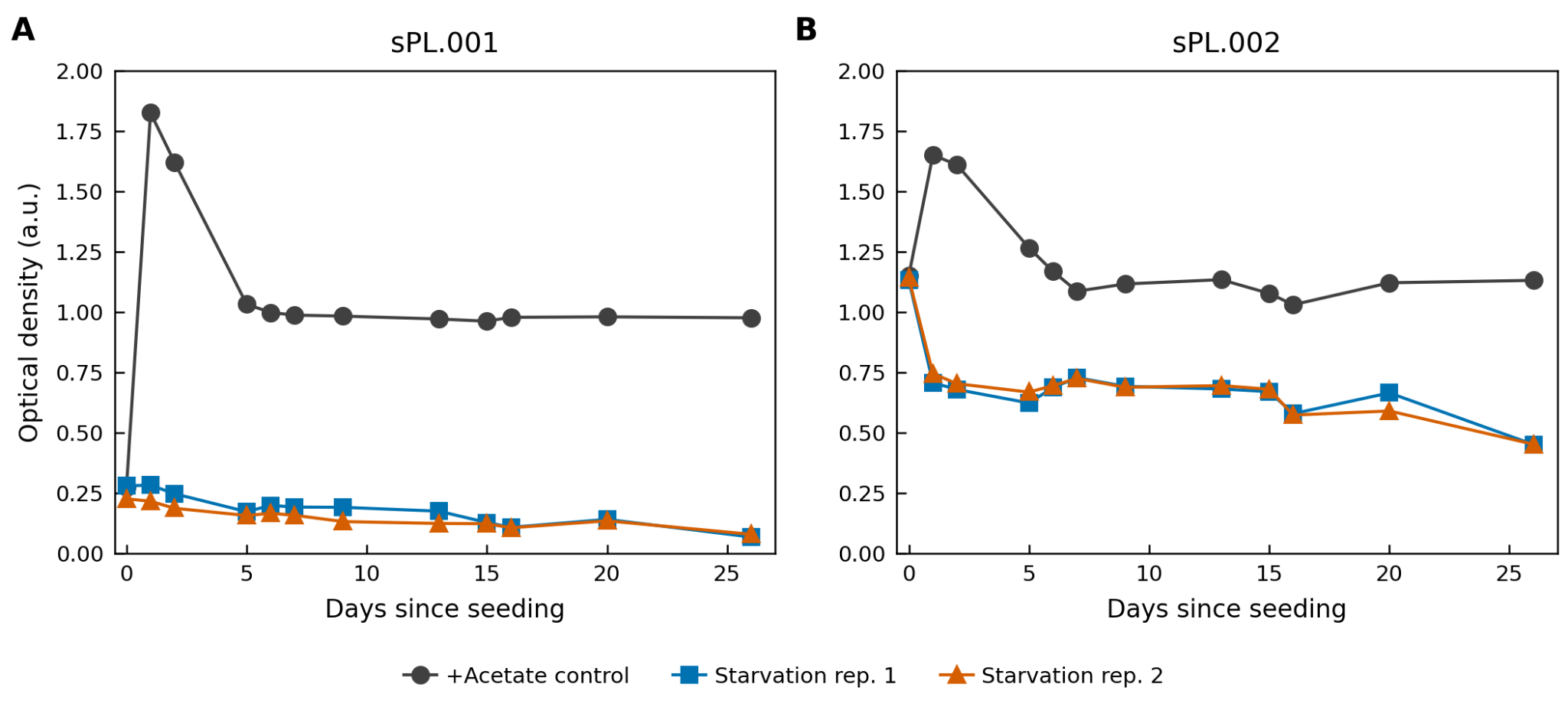}{S2}{Starvation propagation by optical density.}{OD\textsubscript{600nm} over the initial 26 days for sPL.001 (\textbf{A}) and sPL.002 (\textbf{B}). Two replicates (blue squares and orange triangles) and one positive control (black circles) are shown. Starvation conditions declined monotonically. OD readings ended at day 26; CFU data (\textbf{Figure S1}) extended to day 69. Blank subtraction used a sterile 10X DMM reference of OD 0.159.}

\clearpage
\subsection{Low temperature assays}

A single freeze-thaw at -80°C reduced viability only moderately, while
boiling was near-totally lethal, with sPL.001 falling below the 40 cells
mL$^{\mhyphen 1}$ MPN assay detection floor. A single low-level positive result in the
sPL.002 dilution series was judged contamination based on culture appearance, and
both strains are reported at the shared bound.

OD declined steadily in both cold temperature samples of both strains while the
room-temperature controls held near OD 1.0 or above. Viable counts in the cold
temperatures samples declined monotonically from day 0 and remained above the
spot-assay floor throughout.

\piofigurelabeledinline{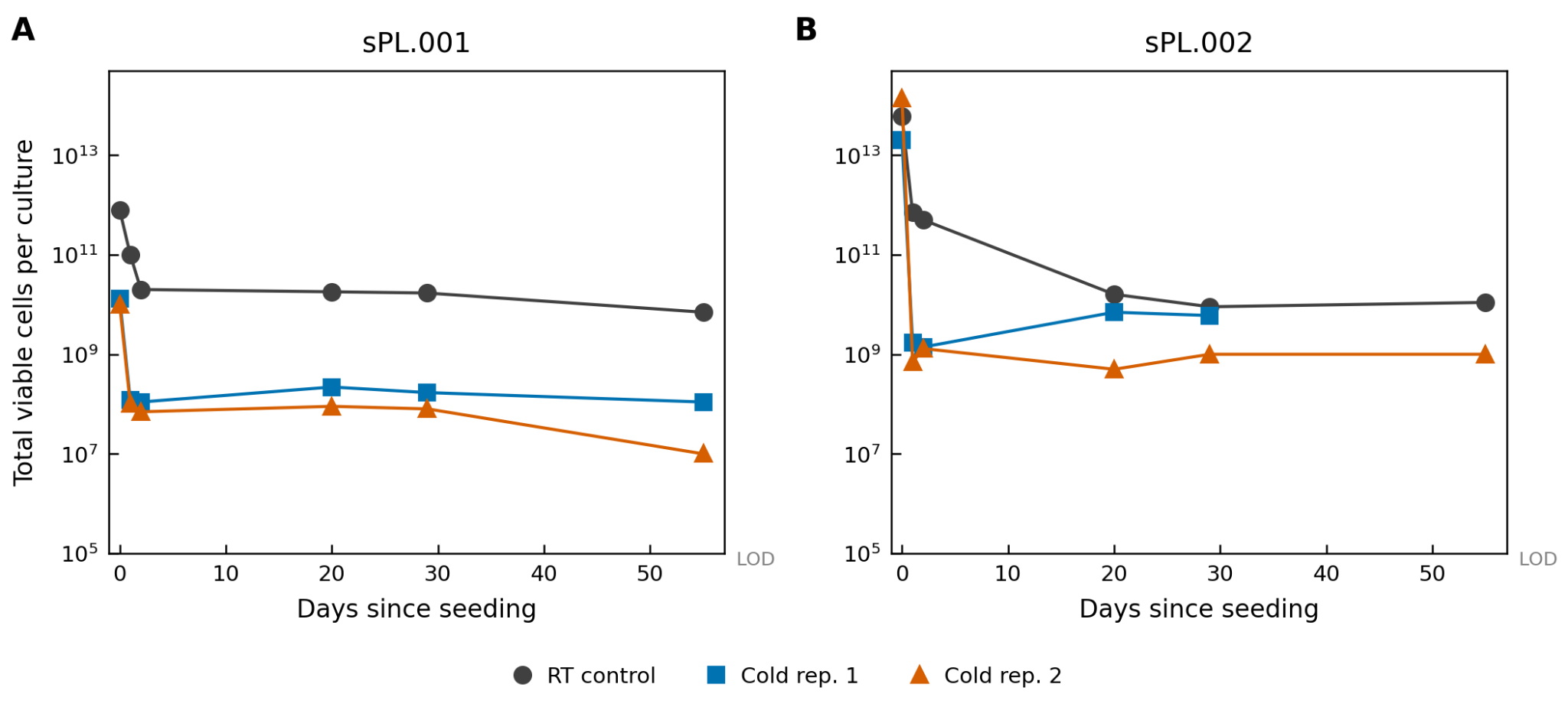}{S3}{Cold propagation (-20°C) through spot plating.}{Total viable cells per culture (log\textsubscript{10} scale) over 55 days for sPL.001 (\textbf{A}) and sPL.002 (\textbf{B}). Two replicates (blue squares and orange triangles) and one positive control (black circles) are shown. sPL.002 cold rep.\ 2 failed to plate at the terminal timepoint and has no day-55 data point. Viable counts in both cold samples declined monotonically and remained above the LOD throughout, with no evidence of propagation.}

\piofigurelabeledinline{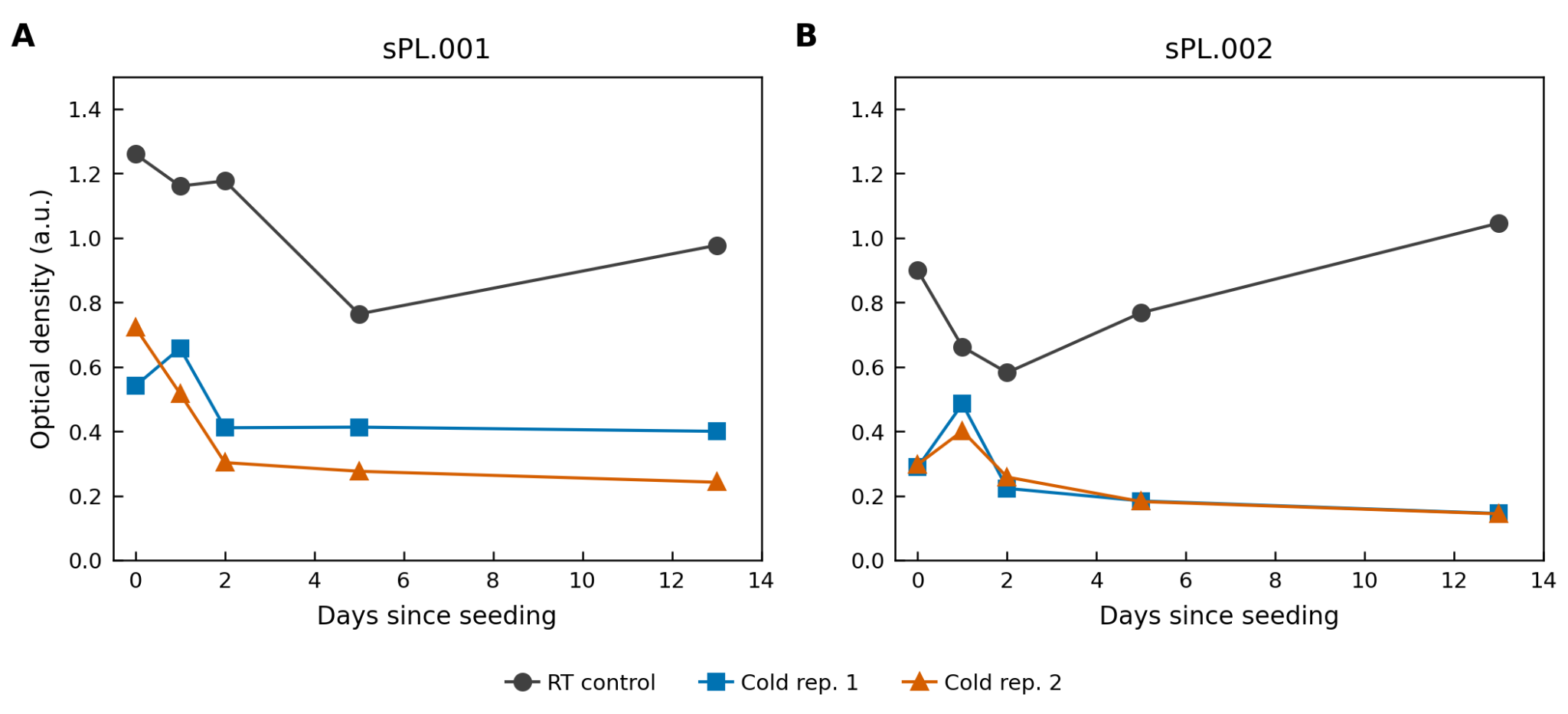}{S4}{Cold propagation (-20°C) through optical density.}{OD (a.u.) over 13 days for sPL.001 (\textbf{A}) and sPL.002 (\textbf{B}). Two replicates (blue squares and orange triangles) and one positive control (black circles) are shown. OD in the cold replicates declined steadily while room-temperature controls held near or above OD 1.0. OD readings ended at day 13 of the 55-day experiment; CFU data (\textbf{Figure S3}) extended to day 55. OD is not directly comparable between RT and -20°C conditions because the cold medium had a different solute composition and no matched blank was used.}

\clearpage
\subsection{Low water activity assays}

In the 56-day propagation experiment, the no-glycerol positive controls grew over
the first two days. sPL.001 from OD 1.214 to 1.926 and sPL.002 from 0.850 to 1.433,
before declining as nutrients were exhausted. Both low-$a_w$ conditions fell below
the spot-plating floor of $\sim 10^{\mhyphen 5}$ of the seeded population within one day and
never recovered.

Terminal whole-culture recovery at day 56 for the low $a_w$ conditions showed zero
recoverable colonies for sPL.002, and a single colony from one of the sPL.001
conditions. 16S sequencing showed it to be a \textit{Micrococcus luteus}
contaminant, likely introduced by the dilution buffer used in the assay.

The sPL.002 positive control contained uncountable thousands of colonies, five of
which were tested with 16S sequencing and validated as sPL.002. However, the
sPL.001 positive control plate resulted in a complex mixture of growth, indicating
contamination. 16S sequencing indicated \textit{Microbacterium luteolum}, as the
primary contaminant with likely fungal accompaniment. However, this is a nearly
identical positive control to the starvation experiment, in which viable sPL.001
was observed after 69 days of growth. The absence of either strain of interest from
every low-water-activity culture is therefore reported as a LOD rather than as a
failure to recover, and the survival and propagation values for this condition are
identical.

\piofigurelabeledinline{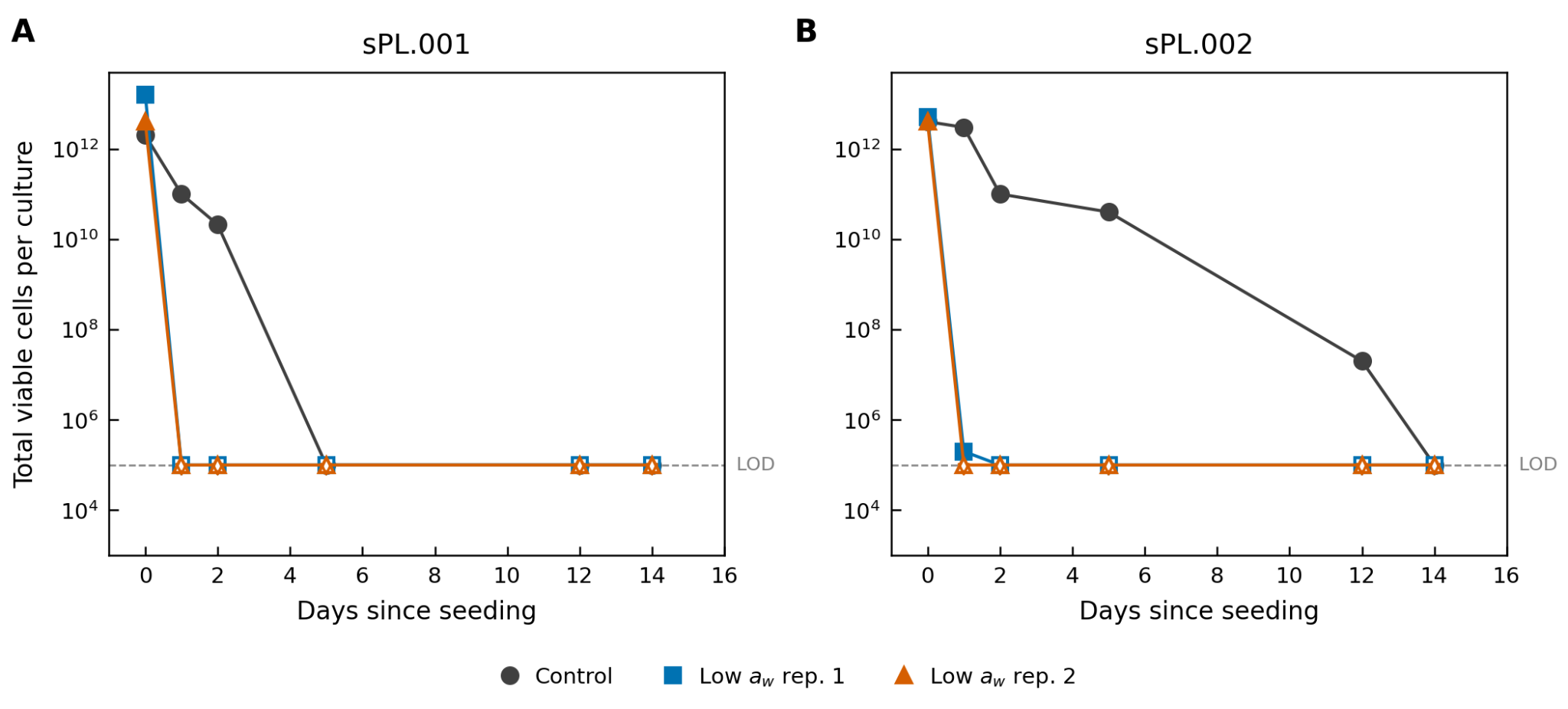}{S5}{Low-water-activity propagation ($a_w$ 0.685) through spot plating.}{Total viable cells per culture (log\textsubscript{10} scale) over 14 days of spot-plating coverage for sPL.001 (\textbf{A}) and sPL.002 (\textbf{B}). Two replicates (blue squares and orange triangles) and one positive control (black circles) are shown. Dashed line indicates the LOD ($1\times10^5$ total CFU per culture). All glycerol condition values at the LOD indicate measurements where no cells were counted.}

\piofigurelabeledinline{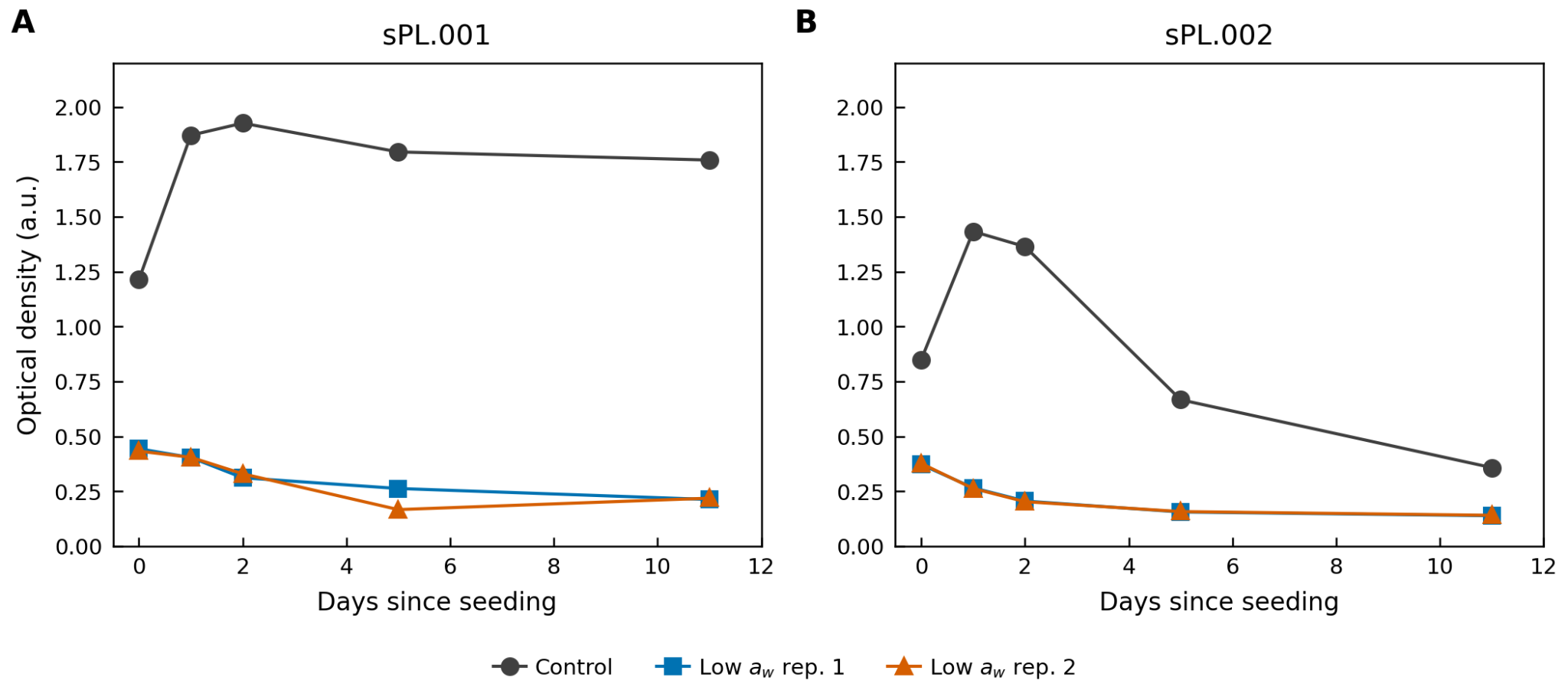}{S6}{Low-water-activity propagation ($a_w$ 0.685) through optical density.}{OD (a.u.) over 11 days for sPL.001 (\textbf{A}) and sPL.002 (\textbf{B}). Two replicates (blue squares and orange triangles) and one positive control (black circles) are shown. OD in glycerol medium is not quantitatively comparable to OD in plain DMM because no glycerol-matched blank was used; the glycerol conditions read 2-3$\times$ lower than controls at day 0 despite equivalent seeding, likely a refractive-index artifact of the high-glycerol matrix rather than immediate cell lysis. OD readings ended at day 11 of the 56-day experiment.}

\end{document}